\documentclass[preprint,aps,groupedaddress,prb]{revtex4}
\usepackage{graphicx}

\begin{document}

\title{Density Matrix Renormalization Group study on incommensurate quantum Frenkel-Kontorova model}
\author{B. Hu$^{1,2}$ and J. X. Wang$^{1}$}

\begin{abstract}
By using the density matrix renormalization group (DMRG) technique,
the incommensurate quantum Frenkel-Kontorova model is investigated numerically. It is found
that when the quantum fluctuation is strong enough, the \emph{g}-function featured by a saw-tooth map in the depinned state will show a different kind of behavior, similar to a standard map, but with reduced magnitude. The related position correlations are studied in details, which leads to a potentially interesting application to the recently well-explored phase transitions in cold atoms loaded in optical lattices.

$PACS:$ 64.60.Ak, 64.70.Rh, 05.10.Cc, 05.45.-a
\end{abstract}

\affiliation{$^{1}$Department of Physics, Centre for Nonlinear Studies, and The Beijing-Hong 
Kong-Singapore Joint Centre for Nonlinear and Complex Systems (Hong Kong), 
Hong Kong Baptist University, Kowloon Tong, Hong Kong, China} 

\affiliation{$^{2}$Department of Physics, University of Houston, Houston, Texas
77204-5005 }

\maketitle

\section{Introduction}
As a generic model to demonstrate the fractal \textquotedblleft devil's
staircase\textquotedblright, 1D Frenkel Kontorova (FK) model has been
investigated extensively in the field of classical nonlinear physics ever
since 1970's. \cite{1} This model describes the dynamics of a chain of particles connected
by springs in the presence of a sinusoidal external potential. Its most
essential feature is the involvement of two competing length scales. One is the average distance between the particles and the other is the spatial period of the external potential. This characteristics can be demonstrated in various physical systems, such as crystal dislocations,\cite{2} epitaxial monolayers on the crystal surface,\cite{3} charge density waves \cite{4} and dry frictions \cite{5} \emph{et al}. Hence studying FK model can lead to the
understanding of many universal properties in those realistic systems.
For example, in the classical regime, when the ratio of two length scales
are rational, i.e. in the commensurate case, any magnitude of external
potential will take the particles into pinned state. But for irrational
ratio, i.e. in the incommensurate case, only when the external potential is
above a threshold can the ground state stay in the pinned state. This later phenomenon is closed related to the breakup of KAM curves into cantorus since the equilibrium positions of the particles are described by the well known Chirikov standard map.\cite{6} This is the so-called
Transition of Breaking Analyticity (TBA) pioneered by Aubry \emph{et al}.\cite{7,8,9}

So far, the classical properties of the 1D FK model have been well understood. As a further step, we need to know how quantum effect will modify those classical results. The research along this line will not only bring us into the topic of the classical-quantum correspondence, but also take us into a fascinating quantum world, where no classical correspondence exists. Unfortunately, until now, there is not much work about it because of the complexes related to the notorious quantum many-body problems. The first work in this direction appeared in 1989,\cite{10} in which by using the Feynman path-integral quantization scheme and the Metropolis algorithm, Borgonovi \emph{et al}. has shown the transition of the \emph{g}-function from a standard map to a saw-tooth one for incommensurate quantum FK model. Similar results have also been reported by the mean-field theory\cite{11} and the squeezed state method.\cite{12,13,14} Recently Krajewski \emph{et al}. has studied the commensurate quantum FK model\cite{15} with the algorithm of path-integral molecular dynamics.

In this paper, instead of making use of the known properties of the classical theory as a starting point for the investigations, we attemp to study the quantum FK model directly. The employed technique is the density matrix renormalization group (DMRG) method.\cite{16} Since its appearance in 1992,\cite{17} through more than one decade of developments,\cite{18} DMRG has now become the most powerful and efficient numerical method to solve one-dimensional quantum many-particle systems. A lot of techniques for the efficiency improvement have been put forward. The details of how to apply them upon quantum FK model will be presented in Section III. 

Besides the theoretical interest, another motivation for us to carry out this work is the rapid experimental development in trapping and controlling Bose-Einstein condensates by the optical lattices.\cite{19,Greiner,Paredes,20} Especially, since the seminal experiment by Greiner \emph{et al}. observing the phase transitions in cold atoms from superfluid (SF) to Mott Insulator (MI),\cite{19} the progresses in this field have been explosive. For the moment, one-, two- and three- dimensional quantum systems can be readily created in experiments.\cite{Koehl}    Theoretically, to describe the underlying physics of these experiments, the Bose-Hubbard model and its various extended versions are widely used.\cite{21,22,23} But the intrinsic tight-binding approximations often require very strong external potentials. On the other hand, when the external potential is very weak, the quantum sine-Gordon model\cite{Coleman} might be an optional model so long as the continuum approximation is applicable. Between the two limiting situations, the minimal model should be quantum FK model, which is actually a discrete version of the sine-Gordon model; hence it goes beyond the limit of continuum or hydrodynamic approximations. It is hoped that our work could not only shed some light on the interpretation of the present on-going experiments, but also help to provide some new insights in modeling and designing the future experiments in wider scope.     

In the following section, the FK model is presented and transformed to the
second-quantized form suitable for the realization of DMRG algorithm. Section 3 explains in detail the DMRG procedures used for quantum FK model. Section 4 is devoted to the results and discussions. In Section V, some physics underlying our calculations are further explored. Finally we will summarize and conclude the work.

\section{Hamiltonian for quantum FK model}

Fig. 1 shows the schematic diagram of the system we study. There are $N+2$ particles connected by $N$ springs and put in a 1D external sinusoidal potential, which could be regarded as representing the effect of a loading optical potential. The fixed-boundary condition is used with $x_0  = 0 $ and $x_{N + 1}  = \left( {N + 1} \right)a$,  in which $x_i, \left(i=0,\cdot\cdot\cdot,N+1 \right)$, is the coordinate of the $i$th particle and $a$ is the average distance between neighbouring particles. The Hamiltonian can be expressed as,
\begin{equation}
\hat H = \sum\limits_{i = 1}^N {\left[ { - \frac{{\hbar ^2 }}{{2m}}\frac{{\partial ^2 }}{{\partial x_i^2 }} + \frac{\gamma }{2}(\hat x_{i + 1}  - \hat x_i  - a)^2  - V\cos (q_0 \hat x_i )} \right]},
\end{equation}
where $q_0$ is the wave number of the external potential, $\gamma$ the force constant of the springs and $V$ the magnitude of the external potential. By introducing the following dimensionless parameters,
\begin{equation}
\hat X_i  = q_0 \hat x_i {\rm{,}}\text{ } \hat U_i  = \hat X_i  - i\mu ,\text{ }{\rm{ }} \hat H' = \frac{{q_0^2 }}{\gamma }\hat H{\rm{,}} \text{ }\mu  = q_0 a,\text{ }{\rm{ }} K = \frac{{Vq_0^2 }}{\gamma },\text{ }{\rm{ }} \tilde \hbar  = \frac{{q_0^2 }}{{\sqrt {m\gamma } }}\hbar, 
\end{equation}
we can rewrite Eq. (1) as,
\begin{equation}
\hat H' = \sum\limits_{i = 1}^N {\left[ { - \frac{{\tilde \hbar ^2 }}{2}\frac{{\partial ^2 }}{{\partial U_i^2 }} + \frac{1}{2}(\hat U_{i + 1}  - \hat U_i )^2  - K\cos (\hat U_i  + i\mu )} \right]}. 
\end{equation}
But Eq. (3) is unwieldly for the realization of DMRG technique. It is necessary to recast it into a second-quantized form. For this purpose, the following substitution is employed,
\begin{equation}
\hat U_i  = \frac{1}{{\sqrt 2 }}\frac{{\sqrt {\tilde \hbar } }}{{\sqrt[4]{2}}}\left( {\hat b_j^{\rm{\dag }}  + \hat b_i } \right),{\rm{  - }}i\frac{\partial }{{\partial U_i }} = i\frac{1}{{\sqrt 2 }}\left( {\hat b_j^{\rm{\dag }}  - \hat b_i } \right),
\end{equation}
in which $\hat b_i^{},\text{ }\hat b_i^{\rm{\dag }}$ are annihilation and creation operators satisfying the boson communication relationship, i.e., ${\rm{ }}\left[ {\hat b_i^{} ,\hat b_j^{\rm{\dag }} } \right] = \delta _{ij}$. Then the Hamiltonian in Eq. (1) is finally reduced to,
\begin{equation}
\begin{array}{l}
 \hat H' = \hat H_0 ' + \hat V_I ', \\ 
 \hat H_0 ' = \sqrt 2 {\rm{ }}\tilde \hbar {\rm{ }}\sum\limits_{i = 1}^N {\left\{ {(\hat b_i^{\rm{\dag }} \hat b_i  + \frac{1}{2}) - \frac{K}{{\sqrt 2 {\rm{ }}\tilde \hbar }}\cos \left[ {\frac{1}{{\sqrt 2 }}\frac{{\sqrt {\tilde \hbar } }}{{\sqrt[4]{2}}}(\hat b_i^{\rm{\dag }}  + \hat b_i ) + i\mu } \right]} \right\}},  \\ 
 \hat V_I ' =  - \frac{{\sqrt 2 {\rm{ }}\tilde \hbar }}{4}\sum\limits_{i = 1}^N {(\hat b_i^{\rm{\dag }}  + \hat b_i )(} \hat b_{i + 1}^{\rm{\dag }}  + \hat b_{i + 1} ). \\ 
 \end{array}
\end{equation}
The above Hamiltonian $\hat H'$ looks quite similar in appearance to that of a lattice system with bosons tight-bounded to the local potentials and only having the nearest-neighbor \textquotedblleft hopping\textquotedblright \text{ }terms. But the physics is quite different. We will come back to this problem later in this paper.
 
By examining Eq. (5), it can be easily seen that there are three independent tunable parameters. The first one is the scaled plank constant $\tilde \hbar$, which is proportional to the square root of the particle kinetic energy, $E_k \sim \hbar^2 q_0^2 /m$, in the unit of the particle-particle interaction energy $\gamma /q_0^2 $, which could be estimated by assuming the particles to occupy the lowest energy level in a 1D potential with a range around $1/q_0$. $\tilde \hbar$ embodies the magnitude of the quantum influence. Experimentally $\tilde \hbar$ can be varied by changing the effective mass $m$ of the particle or the effective strength of the particle-particle interactions. 

The second free parameter is $K$, which measures the strength of the external potential in the unit of $\gamma /q_0^2 $. This is a parameter most easily accessible in experiments since it is directly related to the strength of the loading optical lasers. Generally speaking, the above two parameters are of competing effect, namely, $\tilde \hbar$ tends to delocalize the particles while $K$ to localize them, as is easily understandable and will also be seen from our following calculations. 

The last parameter in Eq. (5) is $\mu  = q_0 a$, which denotes the commensurability property of the system. Because the commensurate-incommensurate phase transition\cite{bak,bak1} is not the focus of this paper. $\mu/2\pi$ is taken to be a fixed irrational number $\left( {\sqrt 5  - 2} \right)/2$ in our work, as conventionally done in literature. It is already well-known that, in classical regime with $\tilde \hbar=0$,  if $\mu/2\pi$ is a rational number, the system will be in a pinned state for an arbitrary external potential. In the incommensurate case, the system can only be in a pinned state when $K$ is larger than a critical value $K_c$; otherwise it will be in a sliding state. But once we take quantum effect into considerations, the above scenario will be greatly modified. Before going to the details, we will first describe the DMRG procedures used for our calculations in the next section.

\section{DMRG procedure}

The essential steps of DMRG method are to form a superblock consisted of two parts denoted by $S$ and $E$. Then, from the targeted state, for example, the ground state $\left| \psi  \right\rangle$ of the superblock, one can obtain the reduced density matrix of $S$ by tracing out the component related to $E$, i.e., $\hat \rho _s  = Tr_E \left| \psi  \right\rangle \left\langle \psi  \right|$. Finally by keeping only $m$ eigenstates of $\hat \rho _s$ with the largest eigenvalues, all the operators related to $S$ will be rebuilt in the space of reduced dimension $m$. It can be proved that the ground states of the rebuilt $S$ has a maximum overlap with the true ground state when $S$ is embedded in the superblock. The error is proportional to the sum of the residue eigenvalues thrown away in the rebuilding processes. Due to the exponential decrease of the eigenvalues from $\rho _s$ for most of the physical systems, very accurate results can still be obtained though working in a much restricted Hilbert space. 

Schematically, the ideas behind DMRG can be explained by Fig. 2. The system $S$ is composed of left block $L$ and left single site $C_{L}$. The environment $E$ is composed of right block $R$ and right single site $C_{R}$. It is assumed that the Hilbert space of $L$ and $R$ have already been truncated to have dimension $m$ with bases written as $\left| \phi _i^L\right\rangle$ and $\left| \phi _i^R\right\rangle, \left( {i = 1,\ldots,m} \right)$, respectively. Initially, the dimension for the full Hilbert space of $S$ is $m^{2}$. Our aim is to seek $m$ most effective bases to truncate this full Hilbert space. DMRG technique has provided us an optimal way to realize this aim. For example, we express the targeted state of the superblock as $\left| \psi  \right\rangle  = \sum\limits_{i,j,k,l = 1}^m {A_{ijkl} \left| {\phi _i^L } \right\rangle \left| {\phi _j^{C_L } } \right\rangle \left| {\phi _k^{C_R } } \right\rangle \left| {\phi _L^R } \right\rangle }$, in which $\left| \phi _j^{C_L } \right\rangle$ and $\left| \phi _k^{C_R }\right\rangle$ are the bases of $C_{L}$ and $C_{R}$. Then the most effective states to represent $S$ is the eigenvectors of the reduced density matrix $\hat \rho _s$ with largest magnitude of eigenvalues. Here $\hat \rho _s  = \sum\limits_{i,j,k,l,r,s = 1}^M {A_{ijrs} A_{rskl}^* \left| {\phi _i^L } \right\rangle } \left| {\phi _j^{C_L } } \right\rangle \left\langle {\phi _k^{C_L } } \right|\left\langle {\phi _l^L } \right|$. Once the Hilbert space for $S$ is truncated in this way, one can add one more single site into $S$ to make a bigger system and then follow the same procedure to make the truncations. Thus step by step, the system size becomes bigger and bigger, while the dimension for the effective Hilbert space keeps to be $m$. This is the so-called infinite algorithm in DMRG language. For finite system of fixed number of particles, as the size of $S$ grows bigger, $E$ will correspondingly becomes smaller. When $S$ reaches the end of the whole system, we will change the role of $S$ and $E$. The truncation procedures are then carried out in the reverse direction along the chain until we get to the other end. This kind of forward and backward "sweeping" processes can be repeated many times. It is the so-called finite algorithm, which is critical to obtain convergent results in DMRG calculations.

For the Hamiltonian expressed by Eq. (5), because the infinite dimension of the local Hibert space, we will not only need to iteratively decimate the space of the growing quantum system, but also should make a systematic truncation over the local Hibert space. The detailed procedures are summarized as follows.

\begin{enumerate}
\item The Hamiltonian in the local Hilbert space is solved numerically for each single particle with the dimension truncated to be $M$,

\begin{equation}
\hat H'_{0i}  = \sqrt 2 {\rm{ }}\tilde \hbar \left\{ {(\hat b_i^{\rm{\dag }} \hat b_i  + \frac{1}{2}) - \frac{K}{{\sqrt 2 {\rm{ }}\tilde \hbar }}\cos \left[ {\frac{1}{{\sqrt 2 }}\frac{{\sqrt {\tilde \hbar } }}{{\sqrt[4]{2}}}(\hat b_i^{\rm{\dag }}  + \hat b_i ) + i\mu } \right]} \right\},
\end{equation}
in which the local basis set for $i$th particle in the Fock space is expressed as $\left| r \right\rangle _i \left( {r = 1,\ldots,M} \right)$. In coordinate representation, 
\begin{equation}
\left\langle x \right.\left| r \right\rangle _i  = \left[ {\frac{{\sqrt 2 }}{{\pi \tilde \hbar 4^r \left( {r{\rm{!}}} \right)^2 }}} \right]^{1/4} H_r \left( {\sqrt {\frac{{\sqrt 2 }}{{\tilde \hbar }}} {\rm{ }}x_i } \right)e^{ - \frac{{x_i^2 }}{{\sqrt 2 \tilde \hbar }}} ,
\end{equation}
with $H_r $ to be the Hermite function of order $r$. Using energy as a weighing measure, only states $\left| \alpha  \right\rangle _i  = \sum\limits_{I = 1}^M {C_{\alpha I} } \left| i \right\rangle _i,\text{ } \left( {\alpha  = 1,\ldots M} \right)$, with the lowest $m$ eigen-energy of Eq. (6) are retained to form the working space in the following steps. It should be mentioned that when solving $\hat H'_{0i}$, the following matrix element is used,
\begin{eqnarray}
\left\langle r \right|_i \cos \left[ {\frac{{\sqrt {\tilde \hbar } }}{{\sqrt 2 \sqrt[4]{2}}}\left( {\hat b_i^\dag   + \hat b_j } \right)} \right]\left| s \right\rangle _i  = \frac{{\sqrt {r{\rm{!}}s{\rm{!}}} }}{{2^{\left( {r + s} \right)/2} }}e^{ - \frac{{\tilde \hbar }}{{4\sqrt 2 }}} \cos \left[ {i\mu  + \frac{\pi }{2}\left( {r + s} \right)} \right]  \times \nonumber \\
\sum\limits_{k = 0}^{\min \left( {r,s} \right)} {\frac{{\left( { - 2} \right)^k  }}{{k{\rm{!}}\left( {r - k} \right){\rm{!}}\left( {s - k} \right){\rm{!}}}}} \left( {\frac{{\sqrt {\tilde \hbar } }}{{\sqrt[4]{2}}}} \right)^{r + s - 2k} ,
\end{eqnarray}
by using,
\begin{equation}
 H_r \left( x \right)H_s \left( x \right) = \sum\limits_{k = 0}^{\min \left( {r,s} \right)} {\frac{{2^k }}{{k{\rm{!}}\left( {r - k} \right){\rm{!}}\left( {s - k} \right){\rm{!}}}}} H_{r + s - 2k} \left( x \right),
\end{equation} 
and 
\begin{equation}
 \frac{1}{{\sqrt {2\pi } }}\int_{ - \infty }^{ + \infty } {e^{ - \left( {z - t} \right)^2 } H_r \left( t \right)dt = 2^r \sqrt \pi  } {\rm{ }}z^n . 
\end{equation}

\item Carry out the finite-size DMRG algorithm with sweeping. The expounding of the corresponding technical details in this step can be found in many papers or books.\cite{16,18} We will not repeat it here. Only some specific techniques adopted in our work are presented as follows.
\begin{enumerate}

\item Conventionally, the superblock in DMRG work is composed of two blocks at the left and right ends together with two single sites in between, which can be schematically expressed as $L \bullet  \bullet R$. If the number of retained states for each block is $m$, the dimension for the superblock would be $m^4$. With $m=20$, it would be an exceedingly intensive calculation to solve a $160000 \times 160000$ matrix repeatedly in the sweeping process. So we will use $L \bullet R$ scheme as done in Ref. 30 since now we only need to solve a $6000 \times 6000$ matrix for the superblock. The amount of calculations are much reduced.

\item During the warmup processes with the infinite DMRG algorithm before sweeping, the bases for the superblock is selected according to the energy cut-off measure.\cite{wei}

\item Once having obtained the solution of the superblock Hamiltonian, the bases of the central site are updated by the eigenstates of the density matrix with the neighboring block parts reduced, i.e., $\hat \rho _c  = Tr_{L,R} \left| \psi  \right\rangle \left\langle \psi  \right|$. If we express the bases of left, central and right parts of the superblock by $
\left| {\phi _i^{L} } \right\rangle ,\left| {\phi _i^{C} } \right\rangle ,\left| {\phi _i^{R} } \right\rangle ,\left( {i = 1,\ldots,m} \right)$, respectively, according to the decreasing order of the eigenvalues from the corresponding reduced density matrix, we can further reduce the dimension of the superblock by keeping only the basis $\left| {\phi _{\left( {i,j,k} \right)}^{super} } \right\rangle  = \left| {\phi _i^{L} } \right\rangle \left| {\phi _j^{C} } \right\rangle \left| {\phi _k^{R} } \right\rangle $ satisfying $i+k+j\le m+2$. Then the superblock matrix size could be decimate again from $m^3$ to $\left[ {\left( {m + 5} \right)\left( {m + 4} \right)\left( {m + 3} \right)/6} \right]$. For example, if $m$ is taken to be $20$, we will have a matrix of size $2300$ instead of $6000$. The effectiveness of this technique has been demonstrated in Ref. 32. 

\end{enumerate}

\item After obtaining a convergent result in Step 2, higher energy levels from the local Hibert space $ \left| \alpha  \right\rangle _i \left( {\alpha  = m + 1,M} \right)$ will be feeded into the systems. To accelerate the calculations, we will feed a group of $m_s$ bases each time. Since the added bases might not be orthogonal to the old ones for each site, all the bases must be orthogonalized and normalized before being submitted to the new round of DMRG sweeping processe in the next step.

\item Continue a new round of DMRG process. But now the central site of the superblock $L \bullet R$ will have a dimension of $m+m_s$. As explained in Step 2(b), the same size-cut technique for the superblock is used once again. It should be mentioned that the procedures adopted in Step 3 and Step 4 is based upon the \textquotedblleft optimized phonon approach\textquotedblright \text{ }, which is often used in applying DMRG algorithm for models involving phonons.\cite{wei,wei1,zhang,fried} But because Eq. (5) is inhomogeneous along the FK chain, the calculations here are much more intensive than those, for example in Holstein model.\cite{zhang} We have to calculate the effective phonon bases for each particle individually since no translation or other symmetries can be used. 

\item Repeat Step 3 by feeding more extra local basis into the systems until we get convergences in both the decimation of the local Hilbert space and the DMRG sweeping. 

\item Goto Step 1 and repeat all the above procedures to ensure the convergence of the results with increasing $M$.

\end{enumerate}

In the following calculations, $M=90, m=10$, and $ m_s=5$ are used. The maximum dimension for the optimized local Hilbert space is $25$, which have been shown to guarantee a numerical convergence. For the test running, we have applied our program upon the exactly solvable model of a pure harmonic chain.\cite{caron} Very accurate results have been obtained. To spare the space, we will not show the details of the test running and the related comparisons with the analytical results here.

\section{Results and discussions}
\subsection{Hull function and \emph{g}-function}
Firstly we check the quantum modifications to the classical hull function and \emph{g}-function, which are defined in quantum version\cite{10} as, 
\begin{equation}
\begin{array}{l}
 F_q (i\mu  + \alpha ) = \bar X_i, \\ 
 g_q {\rm{(\bar X}}_{\rm{i}} {\rm{)  =  }}\frac{{{\rm{\bar X}}_{{\rm{i  + 1}}}  - 2{\rm{\bar X}}_{\rm{i}}  + {\rm{\bar X}}_{{\rm{i  - 1}}} }}{K}, \\ 
 \end{array}
\end{equation}
where $\bar A$ denote the quantum average of the operator $\hat A$ over the ground state $\left| {\psi _0 } \right\rangle$. The results are shown in Fig. 3, from which the quantum smoothening effect can be clearly seen. Firstly, when $\tilde \hbar$ is small, just as expected, there is not much change in the hull and g functions, compared with their classical features, i.e., we have the staircases for the hull function and the discrete points sitting on a sine-function curve for \emph{g}-function. As $\tilde \hbar$ increases until a magnitude around $1\sim 2$, the particles will sample more and more classically-forbidden points along the chain. At the same time, the \emph{g}-function evolves from a form of standard map to a form of saw-tooth map. These results are consistent with those already found in literature.\cite{10,11,12,13,14} From another point of view, they also demonstrates the reliability of our numerical codes. 

The most interesting phenomenon takes place in Fig. 3(c) and 3(d), namely, as $\tilde \hbar$ increases further, the saw-tooth map will become more smoothened and appear in a form similar to a standard map but with a reduced magnitude. Actually, this effect has already shown signs of its existence in the early work from Borgonovi.\cite{bor} But the fuzzy scattered numerical points there deter a clean justification of the obvervance. In our work, because DMRG algorithm can delve into very large $\tilde \hbar$, a clear judgement and detailed study are hence possible. In order to understand what is the underlying physics for this phenomenon, we recall the squeezing state theory,\cite{14} which tells us,
\begin{equation}
g_q  \equiv \frac{{\bar X_{i - 1}  - 2\bar X_i  + \bar X_{i + 1} }}{K} \approx e^{ - \frac{{\tilde \hbar }}{2}G_{ij} } \sin \bar X_i, 
\end{equation}
where $G_{ij}$ represents the correlation function in particle coordinates, i.e., 
\begin{equation}
\tilde \hbar G_{ij}  = \left\langle {\psi _0 } \right|{\rm{ }}\left( {X_i  - \bar X_i } \right)\left( {X_j  - \bar X_j } \right){\rm{ }}\left| {\psi _0 } \right\rangle. 
\end{equation}
As we know, the squeezed state approach meets an almost insurmmoutable challenge in calculating $G_{ij}$ directly due to the involvement in seeking the periodic orbits in a $2(N+1)$-dimension map. But the biggest significance of Eq. (12) lies in the fact that it picks up $G_{ij}$ as the most essential physics in understanding the behavior of $g_q$ especially in deep quantum regime. Within DMRG framework, $G_{ij}$ can be derived readily. Fig. 4 presents the contour map of $G_{ij}$ from which three observances are detailed as follows.
\begin{enumerate}
\item When $\tilde \hbar$ is small ($<<1$), the correlations are confined to a quite narrow region along the diagonal line, which means $e^{-\tilde \hbar G_{ij}}\sim 1$. So reasonably the system's behavior is similar to its classical one according to Eq. (12). 
\item In the region with $\tilde \hbar$ around $1\sim 2$, where a saw-tooth map appears in the $g$-function, we can see some isolated \textquotedblleft islands\textquotedblright \text{ } emerged along the diagonal line in $\tilde \hbar G_{ij}$, shown in Fig. 4(b). This means some strong quantum correlations and fluctuations among the individual parts of the chain. This obvious inhomogeneity of $G_{ii}$ with respect to $i$ could be regarded as being responsible for the formation of the saw-tooth map.  
\item As we explore into the deep quantum region with $\tilde \hbar\gg 2$, a regular pattern of $G_{ij}$ can be observed, which is quite similar to what happens in a pure harmonic chain. For comparisons, we present the case with $K=0$ in Fig. 4(e). The particles are now highly correlated and the previous isolated \textquotedblleft islands\textquotedblright \text{ } merge into each other, which results in a nearly uniform distribution of the diagonal quantum fluctuations. It seems that the quantum fluctuations have strongly renormalized the external potential and make it almost irrelevant as $\tilde \hbar$ flows to infinity. Consequently, according to Eq. (12), the homogeneity of $G_{ii}$ will drive the saw-tooth map to the standard-map-like one, but with reduced magnitude.
\end{enumerate}

Now we can produce a physical picture underlying the evolution of $g_q$ with respect $\tilde \hbar$. In the weak-quantum regime, the particles almost have no correlations. In the middle-quantum regime, the particles start to get correlated with each other from some individual parts of the systems, characterized by the \textquotedblleft islands\textquotedblright \text{ } in the countour map of $G_{ij}$. When more and more particles  becomes correlated, those \textquotedblleft islands\textquotedblright \text{ }will get denser and denser until finally they merge into each other and the systems enter a highly-quantum regime. Since the quantum correlations are closed related to the tunnelling effect, the specific pattern of $G_{ij}$ in the middle quantum regime featured by the saw-tooth map could possibly imply the instanton glass phase ever discussed by Zhirov.\cite{24} But because our present algorithm is restricted to the static ground state, it is very difficult to make the direct investigation of this relationship before we can produce the real-time evolution of the systems. Recently the real-time DMRG (RTDMRG) simulation has attracted wide attentions and achieved a lot of progresses.\cite{25} It will be our next task to extend the present program to explore the dynamic properties of quantum FK model.

In the above discussions, we have seen that the particle coordinate fluctuation $\tilde \hbar G_{ii} $ is a very important quantity to characterize the system dynamics under the quantum influences. For brevity, Fig. 5 presents directly the variation of $\tilde \hbar G_{ii} $ with respect to $i$ for different $\tilde \hbar $. The \textquotedblleft islands\textquotedblright \text{ } observed in Fig. 4 now becomes peaks. Correspondingly, the special feature in the contour map of  $\tilde \hbar G_{ii} $ also appears in Fig. 5. We will not repeat it here. The interesting effect needed to be noted in Fig. 5 is that, for $\tilde \hbar=15$, the curve almost coincide with that for a harmonic chain with $K=0$. It tells us that in the deep quantum region, the external potential plays a negligible role for the system behavior, which corroborates once again the reason for the diminishing amplitude in the $g$-function. 

\subsection{Average fixed-distance correlations}
Because there is no translational symmetry in the incommensurate FK systems, the following average fixed-distance correlations are defined,
\begin{equation}
C_m  = \frac{{\sum\limits_{i,j = 1}^N {\tilde \hbar G_{ij} \delta \left( {\left| {i - j} \right| - m} \right)} }}{{\sum\limits_{i,j = 1}^N {\delta \left( {\left| {i - j} \right| - m} \right)} }}.
\end{equation}
Fig. 6 presents the numerical results of $C_m$. For comparison, the results for a harmonic model without external potential is also plotted. From Fig. 6, we can address three points. Firstly, for small $\tilde \hbar$, there are very little correlations between particles, which approaches their classical behavior in the limit with $\tilde \hbar \rightarrow 0$. Secondly, with increasing $\tilde \hbar$, longer-range particle correlations appear. But instead of monotonous decrease over the particle separation as demonstrated by the harmonic model, there appear fluctuations in the curve. This happens just in the parameter range where saw-tooth map effect emerges. Thirdly, as $\tilde \hbar $ goes beyond about $2$, the curves are getting much smoother without any oscillations until finally approaches the  situation without the external potential. But even in the extremely quantum regime, for example $\tilde \hbar=15$, $C_m$ is still smaller than that in the harmonic model except in the short-separation region. This is a bit different from the results shown in Fig. 5, where nice coincidence of the on-site fluctuations is observed. One reason might be that, compared with the accurate calculations of the ground state energy, there is less precision in calculating the correlations with DMRG algorithm and more over DMRG normally under-estimate the particle correlations. . Another reason is that we are here only considering a finite-size chain. When the distance is as large as the system size, it is understandable to expect higher statistical errors for the long-range correlations

To show the extent of similarity between the FK and the harmonic chains with respect to the particle correlations, we introduce the following order parameter,
\begin{equation}
C = \frac{{\sum\limits_{m = 1}^N {C_m^{FK} } }}{{\sum\limits_{m = 1}^N {C_m^{Harmonic} } }},
\end{equation}
which is just the ratio of the sum for the average fixed-distance correlations in FK chain to that in the harmonic one. Fig. 7 shows the results, from which we see much clearly that as $\tilde \hbar$ increases, there is a transition region with $\tilde \hbar$ around $1\sim 2$. As $\tilde \hbar$ is large enough, $C$ will saturate. The reason for $C$ to be smaller than $1$ should be the same as explained in the above paragraph.

\subsection{Quantum delocalization effect}

Besides the above discussed characteristics of the hull and \emph{g}-functions, we have also studied the delocalization effect of the quantum modifications based upon the particle wave functions. To discuss this properties, the following definitions are used,
\begin{equation}
\begin{array}{l}
 P^{(i)} \left( {X_i } \right) = \int {\psi^{*} \left( {X_1 ,\ldots,X_N } \right)\psi \left( {X_1 ,\ldots,X_N } \right)\mathop{\Pi}\limits_{j \ne i} dX_j },  \\ 
 P\left( X \right) = \frac{\sum\limits_{i = 1}^N {P^{(i)} \left( {X_i  = X} \right)}} {\int_{ - \infty }^{ + \infty } {P\left( X \right)dX}}, \\
\end{array}
\end{equation}
where $\psi \left( {X_1 ,...,X_N } \right)$ is the many-particle wave functions of the system, $P^{(i)} \left( {X_i } \right)$ is the probabilities to find the $i$th particle at position $X_i$ and $P\left( X \right)$ is the total probability to find particles between $X$ and $X+dX$. Firstly, let us check the localization effect of the external potential as done in Ref. 14. Fig. 8 shows the numerical results. It can be easily seen that for fixed $\tilde \hbar$, as the external potential becomes deeper, the particles will tend to sit near the bottom of the potential until finally all the particles are localized and no particles can be found around the top of the potential. This is the localizing effect of external potential. By defining an order parameter to describe the probability for particles to sit on the potential top,
\begin{equation}
P_t  = \frac{1}{N}\sum\limits_{i = 1}^N {\int {\left| {\psi \left( X \right)} \right|} ^2 \delta \left[ {X - 2\pi \left( {n + \frac{1}{2}} \right)} \right]dX}, 
\end{equation}
we can find an abrupt jump in the particle localization as shown in Fig. 9. All the results are consistent with those in Hu and Li's paper.\cite{14} As a further work, we continue to explore into the high quantum region by increasing $\tilde \hbar$. Fig. 10  presents the function $P(X)$ by keeping $K$ fixed. It is interesting to note that as $\tilde \hbar$ increases, more and more particles are found away from the potential bottoms. This is understandable since $\tilde \hbar$ represents the zero-point kinetic energy of the particles. Once again, we plot the disorder parameter $P_t$ in Fig. 11, from which we see another kind of abrupt change of $P(t)$ through a transition region with  $\tilde \hbar$ around $1\sim 2$, where the probability to locate the particles on the top of the potential suddenly increases. Compared with the localization effect of the external potential, the delocalization effect increases much slower. To check the size-dependence of the results, we have shown the calculations with different $N$ in Fig. 11, from which no obvious qualitative change over the above results can be seen. So the quantum delocalization effects seem not much dependent upon the system size. Similar conclusions have also been found in calculating other physical quantities. That is why in most of our work, we only confine to $N=54$.

\section{Relevance of quantum FK model to other models}
In the above section, we have shown how the quantum effect can greatly modify the classical behavior of a FK chain. In this section, we will further discuss the physical significance of quantum FK model by exploring its relationship with other models.

As already mentioned in the introduction, quantum FK model is a minimal model to describe a system of cold atoms loaded in an optical lattice between two limiting cases: one is the Bose-Hubbard model with tight-binding approximations and the other is the quantum sine-Gordon model with hydrodynamic approximations. The marking parameter for these models can be represented by $\beta=K/\tilde \hbar$, which is actually the ratio of the external potential to the kinetic energy of the particles. In the following, we will demonstrate how quantum FK Hamiltonian Eq. (5) looks like with $\beta$ flows to $+\infty$ and $0$.

Firstly, as $\beta \rightarrow+\infty$, this situation can be realized by taking $\tilde \hbar\rightarrow 0$ with $K$ fixed or $K\rightarrow+\infty$ with $\tilde \hbar$ fixed. Both options mean that the particles are tight-bounded by the external potential. Hence the particle index $i$ can be taken as the lattice site index. Once this connection is built, we will try to expand $\hat H_0^\prime$ in Eq. (5) with respect to $\tilde \hbar$ by taking $\tilde \hbar\rightarrow 0$. To simplify the notation, $\xi  = \sqrt {\tilde \hbar }/(\sqrt 2 \sqrt[4]{2})$ is used for the expansion. By utilizing the Baker - Campbell - Hausdorff formula $e^{A + B}  = e^A e^B e^{ - \frac{1}{2}\left[ {A,B} \right]}$ under the condition $\left[ {A,\left[ {A,B} \right]} \right] = \left[ {B,\left[ {A,B} \right]} \right] = 0$, we can have
\begin{eqnarray}
\cos \left[ {\xi \left( {\hat{b}_{i}^{\mathrm{\dag }}+\hat{b}_{i}}\right)
+i\mu }\right]  &=&e^{-\frac{{\xi ^{2}}}{2}}\left\{ {\cos \left( {i\mu }%
\right) \sum\limits_{L=0}^{\infty }{\sum\limits_{j=0}^{2L}{\frac{{\left( {-1}%
\right) ^{L}\xi ^{2L}}}{{j\mathrm{!}\left( {2L-j}\right) \mathrm{!}}}}}%
\left( {\hat{b}_{i}^{\dag }}\right) ^{j}\left( {b_{i}^{\dag }}\right) ^{2L-j}%
\newline
}\right.  \nonumber \\
&&{+\sin \left( {i\mu }\right) \sum\limits_{L=0}^{\infty }{%
\sum\limits_{j=0}^{2L+1}{\frac{{\left( {\mathrm{-1}}\right) ^{L+1}\xi ^{2L+1}%
}}{{j\mathrm{!}\left( {2L+1-j}\right) \mathrm{!}}}}}\left( {\hat{b}%
_{i}^{\dag }}\right) ^{j}\left( {b_{i}^{\dag }}\right) ^{2L+1-j}}.
\end{eqnarray}
The complicated factors in the above sum represents different orders of interactions among the local oscillators through the media of the external potential. We will keep only the terms up to the order of $\xi^4$. Further more, with random phase approximations (RPA), only the operator products $\hat n_i $ and $\hat n_i \hat n_i $ are retained. Then Eq. (5) is reduced to,
\begin{equation}
\begin{array}{l}
 \frac{{\hat H{}^\prime }}{{\sqrt 2 {\rm{ }}\tilde \hbar }} = {\rm{ }}\sum\limits_{i = 1}^N {\left\{ {\left[ {\frac{K}{{\rm{4}}}\cos \left( {i\mu } \right) + 1} \right]\hat n_i  - \frac{{K\tilde \hbar }}{{32\sqrt 2 {\rm{ }}}}\cos \left( {i\mu } \right)\hat n_i \left( {\hat n_i  - 1} \right)} \right\}}  \\ 
 {\rm{             }} - \frac{1}{4}\sum\limits_{i = 1}^N {\left( {\hat b_i^{\rm{\dag }}  + \hat b_i } \right)\left( {\hat b_{i + 1}^{\rm{\dag }}  + \hat b_{i + 1} } \right)}, 
 \end{array}
 \end{equation}
which is stimulatingly similar in form to the following Bose-Hubbard model with disorders,
\begin{equation}
\hat H = \sum\limits_{i = 1}^N {\left[ {\left( {\varepsilon _i  + \mu } \right)\hat n_i  + \frac{U}{2}\hat n_i \left( {\hat n_i  - 1} \right)} \right]}  - t\sum\limits_{i = 1}^N {\left( {\hat b_i^{\rm{\dag }} \hat b_{i + 1}  + \hat b_{i + 1}^{\rm{\dag }} \hat b_i } \right)} ,
\end{equation}
where $t$ denotes the hopping interaction, $U$ is the on-site repulsion, $\epsilon _i$ describes the random external potential and $\mu$ is the chemical potential. If we further assume no phase coherence between ${\hat b_i^{\rm{\dag }} }$ and ${\hat b_{i + 1}^{\rm{\dag }} }$, Eq. (19) and Eq. (20) will appear exactly the same except the fixed $U$ in Eq. (20). This implicates that we can use very similar mathematical technique to solve both Hamiltonians. But these similarities are only valid literally since the underlying physics or interpretations are quite different. Eq. (20) is a true tight-binding single-band Hamiltonian. Thus physically, $\hat b_i^{\rm{\dag }} \left| 0 \right\rangle $ represents the creation of a particle on a definite orbit at the $i$-th site. But in Eq. (19), $\hat b_i^{\rm{\dag }} \left| 0 \right\rangle $ denotes driving the $i$-th particle onto a local state, $\left| r \right\rangle _i$, $(r=1,\cdot\cdot\cdot,\infty)$, as shown in Eq. (7). In other words, Eq. (19) is more like an infinite-band model. Because of this basic difference, we cannot map the results concerning ${\hat b_i }$, which is related to ${\hat X_i }$ and $\hat P_i$, directly to the corresponding ones in Eq. (20).

Now the question becomes, besides as a coordinate operator, what ${\hat X_i }$ implicates in terms of the language from Bose-Hubbard model. The motivation to put forward this question is because a lot of work has already done in Bose-Hubbard model. So the answer will surely help us to extract more physics from our model by analogy. To get some hint of answering this question, we turn to another limit of the quantum FK model with $\tilde \hbar \rightarrow \infty$. One obvious limit is the pure harmonic chain. In Section V, we have already seen this effect in the study of the particle coordinate correlations.

As is well-known, if only the low-energy properties are considered, a universal effective harmonic fluid model can be built for any 1D quantum fluid.\cite{haldane} For example, if we assume the Hamiltonian of a 1D bosonic quantum fluid to be,   
\begin{equation}
H = \int {dx\psi ^\dag  (x)\left( { - \frac{{\hbar ^2 }}{{2m}}} \bigtriangledown ^2\right)} \psi (x) + \frac{1}{2}\int\limits_{ - \infty }^{ + \infty } {dxdx'V(x - x') n(x) n(x')},
\end{equation}
in which $\psi(x)$ is the annihilation operator at position $x$ satisfying $\left[ {\psi (x'),\psi ^\dag  (x)} \right] = \delta (x - x')$ and $n(x) = \psi ^\dag  (x)\psi (x)$. Here, it should be noted that, to keep the notation consistent with that in literature, we have dropped the cap over the operator. To avoid confusions, this notation is only used in the following discussions of this section. By using 
\begin{equation}
\begin{array}{l}
 \psi (x) = \sqrt {n(x)} e^{i\phi (x)},  \\ 
 n(x) =  {n_0  + \Pi {\rm{(x)}}},  \\ 
 \end{array}
\end{equation}
with $n_0=N/L$ the average particle density and assuming that the long-wavelength zero-point fluctuations of the density has a wavenumnber $\ll n_0$, $\phi (x)$ and $\Pi {\rm{(x)}}$ can be treated as conjugate canonical fields,

\begin{equation}
\left[ {\phi (x'),\Pi {\rm{(x)}}} \right] =  - i\delta (x - x').
\end{equation}
With these substitutions and approximations, Eq. (21) becomes,\cite{haldane}
\begin{equation}
H = \frac{{\hbar v}}{{2\pi }}\int {dx\left[ {K\left( {\partial _x \phi (x)} \right)^2  + \frac{1}{K}\left( {\pi \prod (x)} \right)^2 } \right]},
\end{equation}
in which $v$ is the sound velocity and $K$ is a dimensionless parameter determines the quasi-long-range order of the bosonic field. When a sinusoidal external potential $V(x)=Vcos(q_{0}x)$ is added, the above Hamiltonian turns into the famous quantum sine-Gordon form,\cite{bue}

\begin{equation}
H = \frac{{\hbar v}}{{2\pi }}\int {dx\left[ {K\left( {\partial _x \phi (x)} \right)^2  + \frac{1}{K}\left( {\pi \Pi (x)} \right)^2 } \right]}  + Vn_0 \int\limits_{ - \infty }^{ + \infty } {dx\cos \left( {q_{0}\phi (x) + Qx} \right)},
\end{equation}
where $Q$ is a parameter measuring the commensurability property of the model. By discretizing Eq. (25), we obtain a FK Hamiltonian,
\begin{equation}
H = \frac{{\hbar v}}{{2\pi }}\sum\limits_l^{} {\left[ {K\left[ {\phi _l  - \phi _{l + 1} } \right]^2  + \frac{1}{K}\left( {\pi \Pi _l } \right)^2 } \right]}  + {\rm{Vn}}_{\rm{0}} \sum\limits_l^{} {\cos (q_{0}\phi _l  + Ql)}. 
\end{equation}
Compared with Eq. (1), it is obvious that $\hat P_i$ has mapped into $\Pi _i$ and $\hat X_i$ into $\phi _i$. By checking Eq. (22) again, we can immediately see that Fig. 4 actually is a display of the phase coherence between different particles of a realistic quantum system described Eq. (21) in low-energy regime. This is a fascinating scenario since $\hat X_i$ and $\hat P_i$ can be readily calculated by DMRG algorithm. Since the phase coherence is a purely quantum effect, we can now provide another interpretation to the physics in Fig. 4 and Fig. 5. Firstly, when $\tilde \hbar$ is quite small, the particles almost have no phase coherence. Then as $\tilde \hbar$ becomes larger, the phase correlation pattern makes the system more like a bose glass, which has been widely discussed related to disordered Bose-Hubbard model.\cite{fish,krau,sca} When $\tilde \hbar$ takes the system into highly quantum regime, all particles are in a phase-coherent state, much like a superfluid state. With these crude qualitative discussions, we are motivated to follow a remarkable way to explore the non-trivial phase diagrams in quantum FK model by investigating the particle coordinate and momentum correlations. This will provide us an interesting and attractive application of our numerical results to the on-going experiments in the field of cold atoms trapped by optical lattices.  Further work along this direction will is still in progress and will be presented in another paper.
\section{Summary}

In summary, for an incommensurate quantum FK model in the supercritical regime, we can divide the effect of the quantum modifications into three categories.
\begin{enumerate}
\item When $\tilde \hbar \ll 1$, little alternations can be observed in the hull and $g-$ functions, compared with their classical behavior. There is negligible phase coherence among the particles.
\item As $\tilde \hbar$ increases further, for example around $1\sim2$ when $K=5$, individual neighbouring particles begin to be highly correlated. During this process, the saw-tooth map will appear to represent the $g$-function.
\item With $\tilde \hbar$ high enough, more neighboring particles join to strengthen the correlations until all the particles are strongly correlated . At this time, the saw-tooth map will be replaced by a map similar to a standard map, but with much smaller amplitude.
\end{enumerate}

On a qualitative basis, we have shown the relationship of the coordinate correlations calculated in our work to the phase coherence properties in a 1D bosonic system. The interesting variations of the correlation pattern with respect to the quantum modifications could have direct relevance to the interference patterns observed in the superfluid/Mott phase transitions in cold atoms loaded in optical lattices.\cite{an} The systematic investigation will be the focus of our following work. 

This work was supported in part by grants from the Hong Kong Research Grants
Council (RGC) and the Hong Kong Baptist University Faculty Research Grant (FRG).

\newpage 

\Large{Figure Captions}
\normalsize

Figure 1 Schematic diagram of FK model with fixed boundary conditions. There are $N+2$ particles connected by springs. The average distance between the neighboring particles is $a$. $x_0$ and $x_{N+1}$ are fixed to be $0$ and $(N+1)a$.

Figure 2 Schematic diagram for a superblock composed of left block $L$, right block $R$, left central site $C_{L}$ and right central site $C_{R}$. $L$ and $C_{L}$ forms the system part $S$ while $R$ and $C_{R}$ forms the environment part $E$.

Figure 3 Variations of the quantum hull function $\bar X_i$ and g functions $g_q$ with respect to increasing $\tilde \hbar$. $K=5, N=56, \mu/2\pi=34/55$. The same set of values for $K, N, \mu/2\pi$ is used for all the other figures except being specially noted. Because the external potential has a period $2\pi$, $i\mu$ and $\bar X_i$ are folded onto the interval $[0,2\pi]$.

Figure 4 Contour map of the coordinate correlations $\tilde \hbar G_{ij}$ between the $i$-th and the $j$-th particles. For comparison, the case with $K=0$ and $\tilde \hbar=5$ is presented in (e).

Figure 5 The coordinate fluctuations $\tilde \hbar G_{ii}$ of the $i$-th particle with respect to different quantum influences under a fixed external potential with $K=5$. As in Fig. 4, the results corresponding to $K=0$ and $\tilde \hbar=5$ is also presented for comparisons.

Figure 6 Quantum modifications over the average fixed-distance coordinate correlations of the particles $C_m$ under a fixed external potential with $K=5$. The case with $K=0$ ( in solid line ) is shown for reference. For clarity, (a) and (b) correspond to different range of $\tilde \hbar$. In (a), $\tilde \hbar \le 1$ and in (b), $\tilde \hbar > 1$. 

Figure 7 Variations of the order parameter $C$ with respect to $\tilde \hbar$.

Figure 8 Probability distribution $P(X)$ under different strengths of the external potential with $\tilde \hbar=0.2$. The coordinates have been folded onto the region $X \in \left[ {0,2\pi } \right]$.

Figure 9 Total probability $P_t$ to find the particles on the top of the external potential as a function of $K$ for different $\tilde \hbar$.

Figure 10 Probability distribution $P(X)$ under different strengths of the quantum modifications. As in Figure 7, the coordinates have been folded onto the region $X \in \left[ {0,2\pi } \right]$.

Figure 11 Total probability $P_t$ to find the particles on the top of the external potential as a function of $\tilde \hbar$ for different system size $N$.

\newpage

\begin{figure}[tbp]
\includegraphics[angle=0,clip,viewport=-50 0 700 400,width=\textwidth]{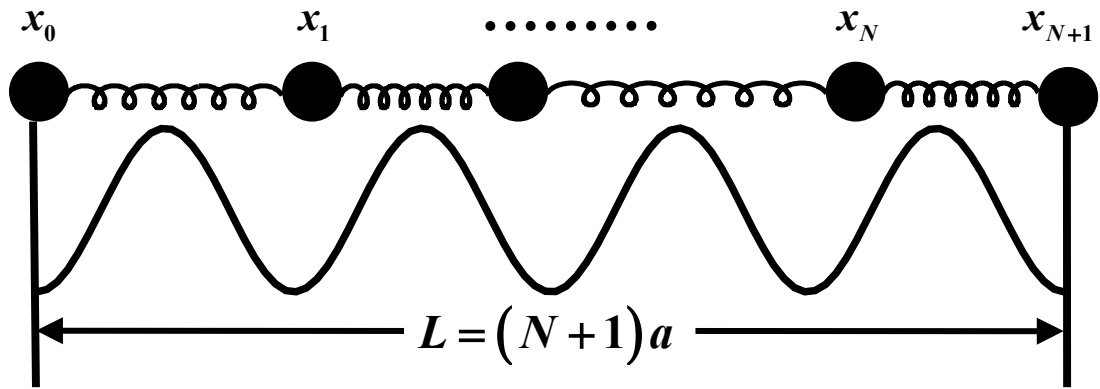} 
\caption{Schematic diagram of FK model with fixed boundary conditions. There
are $N+2$ particles connected by springs. The average distance between the
neighboring particles is $a$. $x_0$ and $x_{N+1}$ are fixed to be $0$ and $%
(N+1)a$. }
\end{figure}
\newpage

\begin{figure}[tbp]
\includegraphics[angle=0,clip,viewport=-50 0 700 400,width=\textwidth]{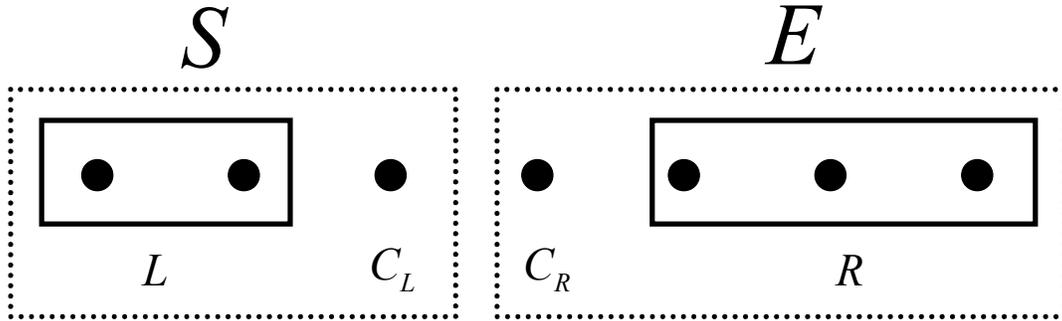}
\caption{Schematic diagram for a superblock composed of left block $L$,
right block $R$, left central site $C_{L}$ and right central site $C_{R}$. $L
$ and $C_{L}$ forms the system part $S$ while $R$ and $C_{R}$ forms the
environment part $E$. }
\end{figure}
\newpage

\begin{figure}[tbp]
\includegraphics[angle=0,clip,viewport=-150 0 700 550,width=1.5\textwidth]{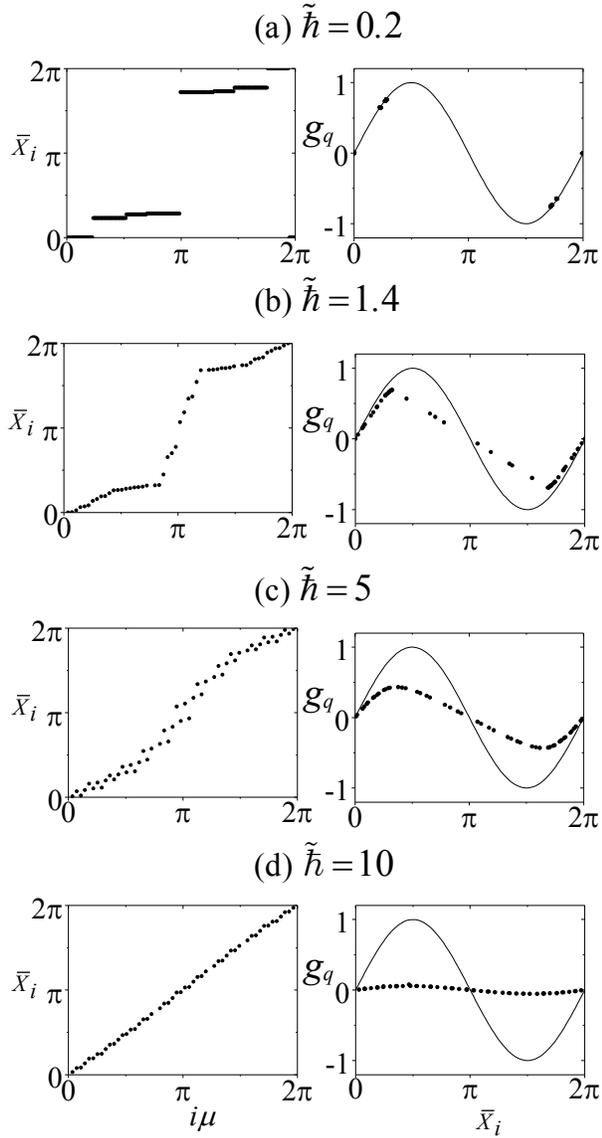} 
\caption{Variations of the quantum hull function $\bar{X}_{i}$ and g
functions $g_{q}$ with respect to increasing $\tilde{\hbar}$. $K=5,N=56,%
\protect\mu /2\protect\pi =34/55$. The same set of values for $K,N,\protect%
\mu /2\protect\pi $ is used for all the other figures except being specially
noted. Because the external potential has a period $2\protect\pi $, $i%
\protect\mu $ and $\bar{X}_{i}$ are folded onto the interval $[0,2\protect%
\pi ]$.}
\end{figure}
\newpage 

\begin{figure}[tbp]
\includegraphics[angle=0,clip,viewport=-150 0 700 700,width=1.5\textwidth]{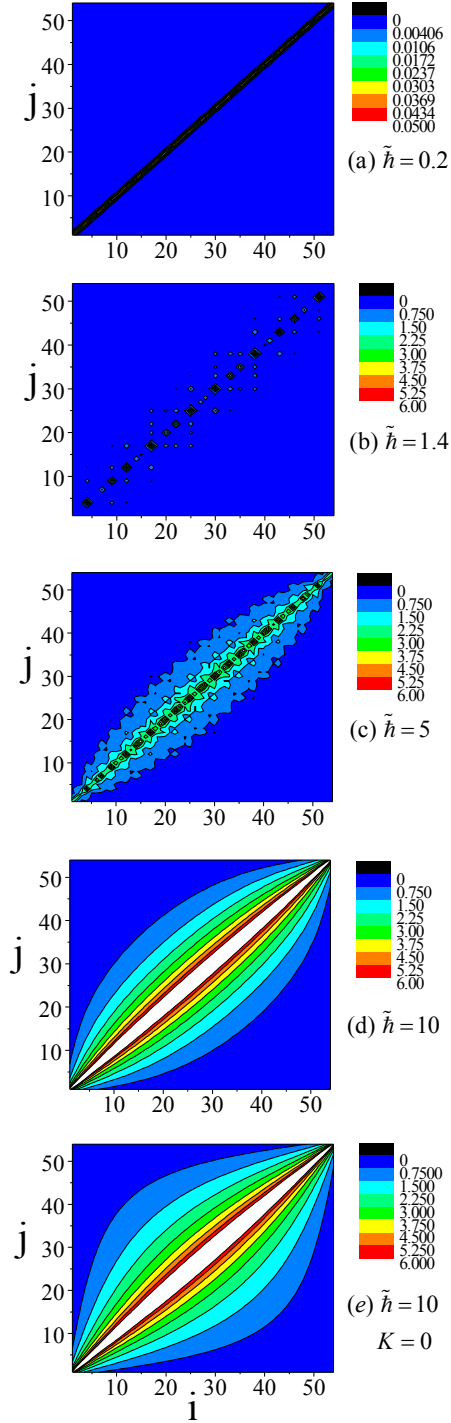} 
\caption{Contour map of the coordinate correlations $\tilde{\hbar}G_{ij}$
between the $i$-th and the $j$-th particles. For comparison, the case with $%
K=0$ and $\tilde{\hbar}=5$ is presented in (e).}
\end{figure}

\newpage 
\begin{figure}[tbp]
\includegraphics[angle=0,clip,viewport=0 0 700 700,width=\textwidth]{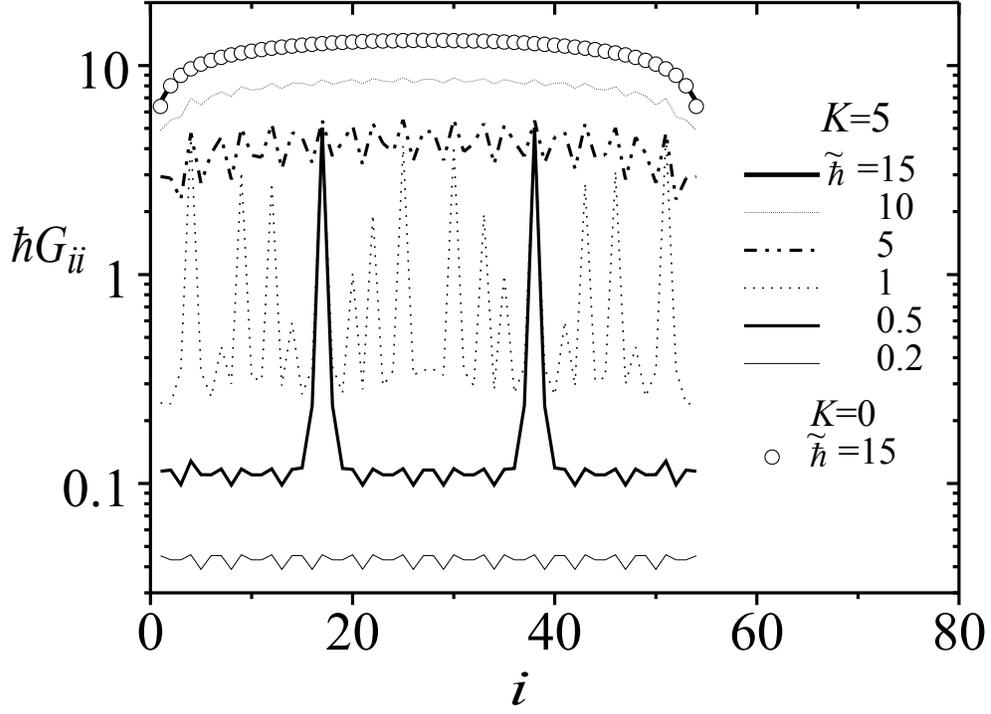} 
\caption{The coordinate fluctuations $\tilde{\hbar}G_{ii}$ of the $i$-th
particle with respect to different quantum influences under a fixed external
potential with $K=5$. As in Fig. 4, the results corresponding to $K=0$ and $%
\tilde{\hbar}=5$ is also presented for comparisons.}
\end{figure}

\newpage 
\begin{figure}[tbp]
\includegraphics[angle=0,clip,viewport=-100 0 700 500,width=1.5\textwidth]{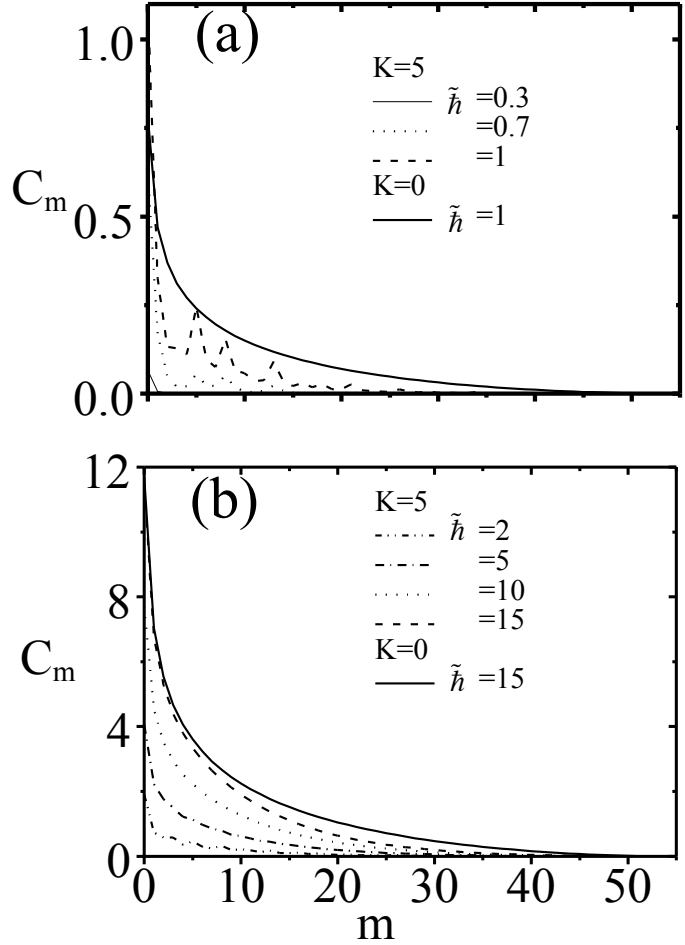} 
\caption{Quantum modifications over the average fixed-distance coordinate
correlations of the particles $C_{m}$ under a fixed external potential with $%
K=5$. The case with $K=0$ ( in solid line ) is shown for reference. For
clarity, (a) and (b) correspond to different range of $\tilde{\hbar}$. In
(a), $\tilde{\hbar}\leq 1$ and in (b), $\tilde{\hbar}>1$. }
\end{figure}

\newpage 
\begin{figure}[tbp]
\includegraphics[angle=90,clip,viewport=400 100 800 800]{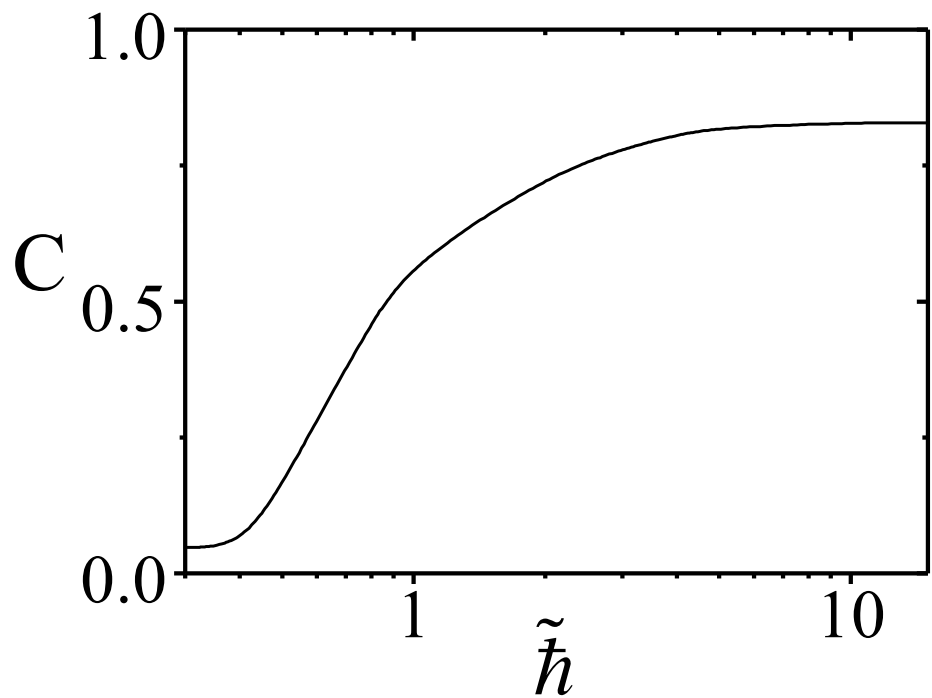} 
\caption{Variations of the order parameter $C$ with respect to $\tilde{\hbar}
$.}
\end{figure}

\newpage 
\begin{figure}[tbp]
\includegraphics[angle=0,clip,viewport=0 0 800 700,width=0.7\textwidth]{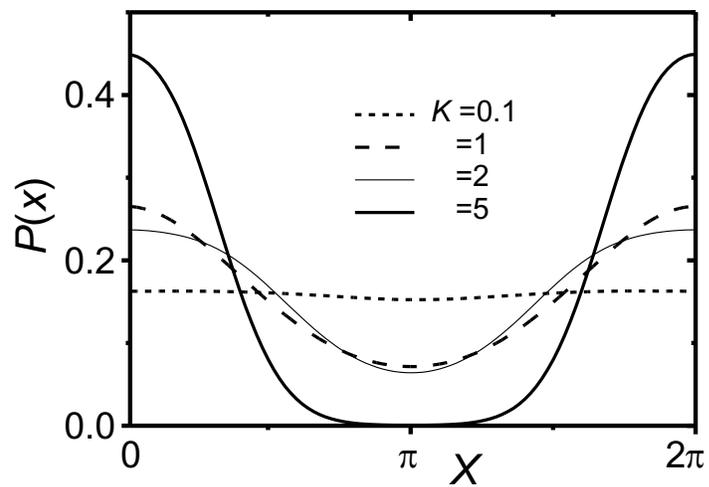} 
\caption{Probability distribution $P(X)$ under different strengths of the
external potential with $\tilde{\hbar}=0.2$. The coordinates have been
folded onto the region $X\in \left[ {0,2\protect\pi }\right] $.}
\end{figure}

\newpage 
\begin{figure}[htbp]
\includegraphics[angle=90,clip,viewport=100 100 500 500]{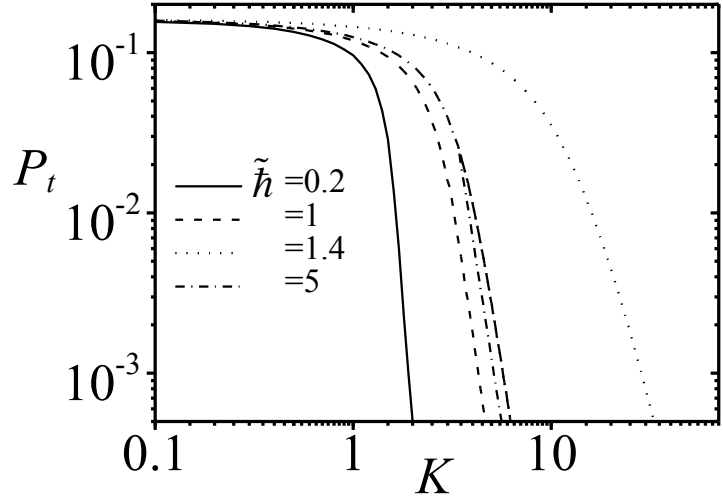} 
\caption{Total probability $P_{t}$ to find the particles on the top of the
external potential as a function of $K$ for different $\tilde{\hbar}$.}
\end{figure}

\newpage 
\begin{figure}[tbp]
\includegraphics[angle=90,clip,viewport=100 100 400 400]{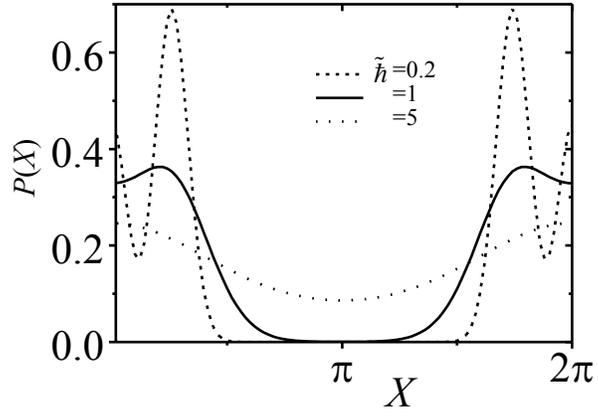} 
\caption{Probability distribution $P(X)$ under different strengths of the
quantum modifications. As in Figure 7, the coordinates have been folded onto
the region $X\in \left[ {0,2\protect\pi }\right] $. }
\end{figure}

\newpage 
\begin{figure}[tbp]
\includegraphics[angle=0,clip,viewport=0 0 800 750,width=0.7\textwidth]{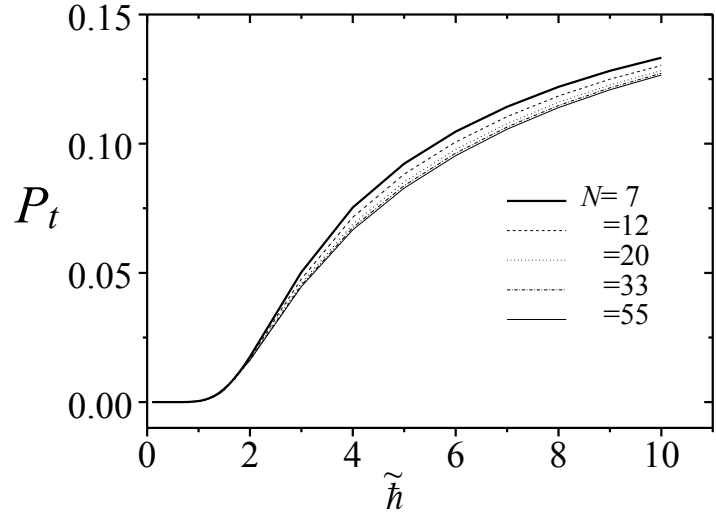} 
\caption{Total probability $P_{t}$ to find the particles on the top of the
external potential as a function of $\tilde{\hbar}$ for different system
size $N$.}
\end{figure}

\end{document}